\documentstyle[emulateapj,apjfonts]{article}
\newcommand{\lsim}{\mathrel{\mathop{\kern 0pt \rlap{\raise.2ex\hbox{$<$}}}
\lower.9ex\hbox{\kern-.190em $\sim$}}}
\newcommand{\gsim}{\mbox{\raisebox{-1.0ex}{$~\stackrel{\textstyle >}
{\textstyle \sim}~$ }}}
\newcommand{\be}{\begin{equation}}
\newcommand{\ee}{\end{equation}}
\newcommand{\ba}{\begin{eqnarray}}
\newcommand{\ea}{\end{eqnarray}}
\newcommand{\nn}{\nonumber}
\newcommand{\gm}{\gamma}

\newcommand{\epf}{\epsilon_f}
\newcommand{\epff}{\epsilon_{f}^{2}}
\received{JUNE 1999}
\slugcomment{To appear in ApJ (Feb 20, 2000), vol. 530 (astro-ph/9906239)}

\lefthead{Suh \& Mathews}
\righthead{Mass-Radius Relation for Magnetic White Dwarfs}

\begin{document}

\title{Mass-Radius Relation for Magnetic White Dwarfs}

\author{In-Saeng Suh$^{1}$ and G. J. Mathews$^{2}$}   

\affil{Center for Astrophysics,
Department of Physics,
University of Notre Dame, 
Notre Dame, Indiana 46556, USA \\
$^{1}$isuh@cygnus.phys.nd.edu; $\;\;$
$^{2}$gmathews@bootes.phys.nd.edu}
\date{\today}

\begin{abstract}
Recently, several white dwarfs with very strong surface magnetic fields have been
observed. In this paper we explore the possibility that such stars could have 
sufficiently strong internal fields to alter their structure.
We obtain a revised  white dwarf mass-radius relation in the presence
of strong internal magnetic fields. We first derive the equation of state for a fully
degenerate ideal electron gas in a magnetic field using an Euler-MacLaurin expansion.
We use this to obtain the mass-radius relation for magnetic $^{4}$He, $^{12}$C, and
$^{56}$Fe white dwarfs of uniform composition.
\end{abstract}


\section{Introduction}

A number of white dwarfs with strong magnetic fields have been discovered 
(\cite{kemp}; \cite{putney}; \cite{schmidt95}; \cite{reimers}) 
and extensively studied (\cite{jordan92}; \cite{angle}; \cite{chanmugam} and references 
therein). Surface magnetic fields ranging from about $10^5$ G 
to $10^9$ G have been detected in about 50 (2$\%$) of the $\approx 2100$ known white
dwarfs (\cite{jordan97} and references therein).
As relics of stellar interiors, the study of the magnetic fields in and around 
degenerate stars should give important information on the role such fields play in
star formation and stellar evolution. However, the origin and evolution of stellar
magnetic fields remains obscure.

As early as Ginzburg (1964) and Woltjer (1964) it was proposed that the magnetic 
flux ($\Phi_B \sim BR^2$) of a star is conserved during its evolution and 
subsequent collapse to form a remnant white dwarf or neutron star. A main sequence star
with radius on the order of $R \sim 10^{11}$ cm and surface magnetic field 
$B \sim 10 - 10^4$ G 
[magnetic A-type stars have typical surface fields $\lsim 10^4$ G 
(\cite{ST})] would thus collapse to form a white dwarf with 
$R \sim 10^9$ cm and $B \sim 10^5 - 10^8$ G, 
or a neutron star with $R \sim 10^6$ cm and $B \sim 10^{11} - 10^{14}$ G.
Indeed, shortly after their discovery (Hewish et al. 1968) pulsars 
were identified as rotating neutron stars
(\cite{gold}) with magnetic fields $B \sim 10^{11} - 10^{13}$ G
consistent with magnetic field amplification by flux conservation. 
In addition, neutron stars with surface magnetic fields exceeding $10^{14}$ G
[so called magnetars] have been recently suggested as the source of soft gamma-ray 
repeaters (\cite{duncan}; \cite{thompson}). 

Moreover, the surface magnetic field of a star does not necessarily reflect the 
internal field (\cite{ruderman}). 
For example, the toroidal fields below the surface of the Sun are
at least on the order of $\sim 10^2$ to $\sim 10^4$ times stronger than the average
surface dipole field strength of $\sim 1$ G (\cite{galloway}). 
Furthermore, at the region of the convective zone, the strength of small scale magnetic
fields could reach a value as high as $7 \times 10^{4}$ G (\cite{chauhan,pulido}).
This would correspond to an interior field strength on the order of $\sim 10^{9}$ to 
$\sim 10^{13}$ G
in a white dwarf, or $\sim 10^{15}$ to $\sim 10^{18}$ G in a neutron
star. [Condensed objects of size $R$ and mass $M$ have an upper limit 
to their field strengths of $B \lsim M R^2 (8 \pi G)^{1/2}$. 
For neutron stars with $R \approx 10$ km and 
$M \approx M_{\odot}$, the limit is $B \lsim \sim 10^{18}$ G (\cite{lerche}).]

Indeed, the existence of white dwarfs with interior magnetic fields as strong as 
$\sim 4 \times 10^{13}$ G is not ruled out with the present uncertainties
in the mass-radius relation (\cite{ST}). 
The present high upper limit on the strength of internal fields in white dwarfs
is obtained by simply setting the magnetic pressure equal to the internal pressure of the
star. However, white dwarfs with internal fields at or around this strength 
could be constrained (\cite{mestel}) by a perceptibly different mass-radius relation.

Although white dwarfs in binaries with well determined masses do not appear
to have surface magnetic fields larger than $\sim 10^5$ G,
internal fields of order $10^{12}$ G could be well hidden below the surface (\cite{angle}).
Newly discovered magnetic degenerate stars, especially those with surface field strengths
near the range of $B \sim 10^9$ G, always show strong circularly and/or
linearly polarized spectral energy distributions (\cite{schmidt99}).
Moreover, these stars reveal unique spectral features (\cite{engelhardt}) due to
quasi-Landau resonances in extremely high magnetic fields of $> 10^9$ G.
 
In this work, we explicitly compute the mass-radius relation of white dwarfs with 
internal magnetic fields. 
Previously, Ostriker \& Hartwick (1968) have estimated effects of interior magnetic
fields by considering a correction in terms of the ratio of magnetic to
gravitational energy. They showed that a relatively small ratio
of magnetic to gravitational energy would be sufficient to explain an observational
discrepancy in the classical mass-radius relation for Sirius B.
However, if white dwarfs could indeed have central magnetic fields as strong as   
$4.4 \times 10^{11} - 4.4 \times 10^{13}$ G, the revised mass-radius relation must be
explicitly determined by taking the magnetic field into account in the equation of state. 
The present work thus expands upon that earlier study by explicitly computing 
the equation of state for a completely degenerate, noninteracting electron gas in a
magnetic field. This equation of state is then applied to the Tolmam-Oppenheimer-Volkoff
(TOV) equation of stellar hydrostatic equilibrium. 

The equation of state in a magnetic field should reduce to a normal
equation of state in the absence of a magnetic field. Therefore, 
we use an Euler-MacLaurin expansion (\cite{kernan}) of the thermodynamic variables
to recover the weak field limit. 
In integrating the TOV equation we simply follow the procedure of
Hamada \& Salpeter (1961) for degenerate matter of uniform composition of $^{4}$He,
$^{12}$C, or $^{56}$Fe. Although $^{4}$He and $^{12}$C white dwarfs are expected 
to have a similar (though not identical) mass-radius relation, 
we explicitly consider each for completeness.

\section{Equation of state for an electron gas in a magnetic field}

The properties of an electron in an external magnetic field have been
studied extensively (\cite{landau}; \cite{JL}; \cite{canuto}; \cite{schwinger}). 
In brief, the energy states of an electron in a
magnetic field are quantized and its properties are modified
accordingly. In order to investigate these effects,
we must first solve the Dirac equation in an external, static, and homogeneous
magnetic field. We make the convenient choice of gauge for the vector potential in 
which a uniform magnetic field $B$ lies along the $z$-axis. 
We then obtain the electron wavefunctions and energy dispersion in a magnetic
field (\cite{JL}).
The dispersion relation for an electron propagating through
a magnetic field is   
\be
E_n = [p^2 c^2+ m^2 c^4+ 2 \hbar c e B n]^{1/2},
\ee  
where $n = j + {1\over 2} + s_z, \, (n=0, 1, \, \ldots$), 
$j$ is the  principal quantum number of the Landau level, 
$s_z =\pm 1/2$ is the electron spin, $e$ is the electron charge, $c$ is the speed 
of light, $\hbar$ is Planck's constant, 
$p \equiv p_z$ is the electron momentum along the $z$-axis, and 
$m$ is the rest mass of the electron.
In Eq. (1) we ignore the anomalous magnetic moment for an electron.

The main modification of an electron in a magnetic field comes from the available
density of states for the electrons (\cite{landau}). 
The electron state density in the absence of a magnetic field, 
\[
\frac{2}{\hbar^3} \int \frac{d^{3}\vec{p}}{(2 \pi)^3},
\]
is replaced with
\[
\frac{2}{\hbar^2 c} \sum_{n=0}^{\infty}[2-\delta_{n \, 0}] 
\int {eB\over(2\pi)^2} dp_z
\]
in a magnetic field. The symbol $\delta_{n \, 0}$ is the Kronecker delta
defined by $\delta_{n \, 0} = 1$ for $n=0$, or $\delta_{n \, 0} = 0$ for $n \ne 0$.
This modification affects the thermodynamic variables for the 
electron gas. 

An isolated white dwarf ultimately cools to zero temperature (fully degenerate) and
the degeneracy pressure supports these stars against 
further gravitational collapse. 
For the most part, this pressure can be described as an ideal
(noninteracting) electron gas plus small corrections.

Let us consider a gas of electrons at zero temperature in a magnetic field.
From Eq. (1) we can define the Fermi energy $E_F$ for an arbitrary Landau level $n$
as
\be
E_{F}^{2} \equiv m^2 c^4 + p_{F}^2 c^2 + 2 \hbar c e B n. 
\ee
Here $p_F$ denotes the Fermi momentum. The number density of electrons 
in a magnetic field is then given by 
\be
n_e  = 2 \frac{\gm}{(2 \pi)^2} \left( \frac{m c}{\hbar} \right)^3
\zeta(\epf, n),
\ee
where
\[
\zeta(\epf, n) = \sum_{n=0}^{n_f} [2 - \delta_{n \, 0}] \;\sqrt{\epff - (1 + 2 \gm n)}.
\]
In the above, $\epf \equiv E_F / m c^2$ and  
$\gamma = B/B_c$ where $B_{c} = m^{2} c^3 /|e| \hbar = 4.414 \times 10^{13}$ G is 
the critical magnetic field at which quantized cyclotron states begin to exist.  
The maximum Landau level $n_f$ for a given Fermi energy $\epf$ and magnetic field 
strength $\gm$
is given by 
\[
n_f \equiv \frac{\epff - 1}{2 \gm} \ge n. 
\]
The pressure of an ideal electron gas in a magnetic field is then
\be 
P_e = 2 \frac{\gm}{4 \pi^2} m c^2 \left( \frac{m c}{\hbar} \right)^3
\; \Phi(\epf,n),
\ee
where
\ba
\Phi(\epf,n) &=& \frac{1}{2} \sum_{n=0}^{n_f} [2 - \delta_{n \, 0}] \;
\Bigg[ \epf \sqrt{\epff - (1 + 2 \gm n)} \nn \\
&~&-(1 + 2 \gm n) \; \mbox{ln}\Bigg( \frac{\epf + \sqrt{\epff -
(1 + 2 \gm n)}}{\sqrt{1 + 2 \gm n}} \Bigg) \Bigg]. \nn
\ea
Similarly, the energy density is 
 \be
{\cal E} (\epf,n) = 2 \frac{\gm}{4 \pi^2} m c^2 \left( \frac{m c}{\hbar} \right)^3
\; \chi(\epf,n),
\ee
where
\ba
\chi(\epf,n) &=& \frac{1}{2} \sum_{n=0}^{n_f} [2 - \delta_{n \, 0}] \;
\Bigg[ \epf \sqrt{\epff - (1 + 2 \gm n)} \nn \\
&~&+(1 + 2 \gm n) \; \mbox{ln}\Bigg( \frac{\epf + \sqrt{\epff - 
(1 + 2 \gm n)}}{\sqrt{1 + 2 \gm n}} \Bigg) \Bigg]. \nn
\ea
From these, we obtain the energy per electron
\be
E_e (\epf,n) = m c^2  \; \frac{\chi(\epf,n)}{\zeta(\epf,n)}.
\ee
  
In order to recover the usual equation of state in the absence of a magnetic field,
we utilize an Euler-MacLaurin expansion of Eqs. (3) - (6) in the weak field limit.
Then, the number density is given by
\be
n_e \simeq \frac{1}{3 \pi^2} \left( \frac{m c}{\hbar} \right)^3
\; \zeta(x),
\ee
where,
\[
\zeta(x) =  x^3 + \gm^2 \frac{1}{4 x} + {\cal O}(\gm^4) + \cdots,
\]
and $x \equiv p_F / mc$ is the relativity parameter.
Note that the electron number density
increases as the magnetic field increases for a given $x$.

The pressure becomes
\be
P_e \simeq \frac{1}{24 \pi^2} m c^2 \left( \frac{m c}{\hbar} \right)^3 
\; \Phi (x),
\ee
where
\[
\Phi (x) = \Phi_{0}(x) + \gm^2 \Phi_{B}(x) + {\cal O}(\gm^4) + \cdots,
\]
\[
\Phi_0 (x)= x \sqrt{x^2 + 1} (2 x^2 - 3) + 3 \, \mbox{ln} (x + \sqrt{x^2 + 1}),
\]
\ba
\Phi_{B}(x)&=&  \frac{\sqrt{x^2+1}}{x} + 2 \, \mbox{ln} (x + \sqrt{x^2 + 1}) \nn \\
&~&- \left( 1 + \frac{1}{x (x + \sqrt{x^2 + 1})} \right). \nn
\ea
Note also that for a physically reasonable value of $x$ the pressure always increases 
as $\gm$ increases.

The energy density can also be written
\be
{\cal E}(x) \simeq  m c^2 \left( \frac{m c}{\hbar} \right)^3  \; \chi(x),
\ee
where
\[
\chi(x) = \chi_0 (x) + \gm^2 \chi_{B} (x) + {\cal O}(\gm^4) + \cdots,
\]
\[
\chi_0 (x) = \frac{1}{8 \pi^2} \Big[ x (2 x^2 + 1) \sqrt{x^2 + 1}
- \mbox{ln} (x + \sqrt{x^2 + 1}) \Big],
\]
\ba
\chi_{B} (x) &=& \frac{1}{24 \pi^2} \Bigg[
1 + \frac{\sqrt{x^2 + 1}}{x} + \frac{1}{x(x+\sqrt{x^2 + 1})} \nn \\   
&~&- 2 \, \mbox{ln} (x + \sqrt{x^2 + 1}) \Bigg]. \nn
\ea

Finally, the energy per electron is given by 
\be
E_e (x) \simeq \frac{3}{8} m c^2  \; \frac{\chi(x)}{\zeta(x)}.
\ee
Here we can see explicitly that as $\gm$ goes to zero, Eqs. (7) - (10) recover exactly
the usual equation of state in the absence of a magnetic field.

\section{Mass-radius relation of magnetic white dwarfs}

The mass-radius relation of white dwarfs was first determined by Chandrasekhar (1939). 
Later Hamada \& Salpeter (1961) obtained numerical models for various core
compositions by considering a fully degenerate configuration at zero temperature. 
The theoretical relationship between the mass and radius of a white dwarf is important
for the interpretation of observational results (see \cite{koester} for a review).
There are several recent studies and observations 
on the mass-radius relation of non-magnetic white dwarfs
(\cite{wood}; \cite{vauclair}; \cite{vennes95}; \cite{provencal}).

In order to obtain the mass-radius relation for magnetic white dwarfs, 
we use Eqs. (7) - (10) for $\gamma \le 1$ ($B \le 4.4 \times 10^{13}$ G) and 
carry out stellar integrations for a uniform composition 
of $^{4}$He, $^{12}$C, and $^{56}$Fe as an illustrative model.
In this uniform model (\cite{hamada}; \cite{fushiki}; \cite{rog})
the total energy $E$ of the plasma consists of a nearly
uniform distribution of degenerate electrons with embedded ions,
\be
E = E_e + E_{C},
\ee
where the first term is the energy of a uniform gas of free electrons and
$E_{C}$ corrects for the classical Coulomb energy.
Although the noninteracting electron gas accounts for the dominant contribution
to the equation of state at high density,
the classical Coulomb correction is significant. 
Other corrections, such as the Thomas-Fermi, exchange, and correlation corrections
give only a very small change in the mass-radius relation of white dwarfs.
Actually the Thomas-Fermi correction in a strong magnetic field is important at low
density (see \cite{rog} and references therein), but as a whole it
gives only a minor effect on the mass-radius relation. Hence, we ignore
these minor effects for the present work.   

Magnetic fields should not alter the spherical symmetry of the constituent atoms.
Indeed, it is, perhaps, a remarkable fact that magnetic fields do not destroy even 
the approximate spherical symmetry of heavy atoms within the relevant range 
(\cite{rog}). 
Thus, we can use the ordinary electrostatic energy (Coulomb energy) per
charge,
\be
E_C / Z = - \frac{9}{5} Z^{2/3} Ry \frac{1}{r_e},
\ee
where $Ry = \frac{1}{2} \alpha^2 m_e c^2$ is the Rydberg energy and from Eq. (7),
\[
r_e = \left(\frac{3 \pi}{8 \gm \zeta(x)} \right)^{1/3} \alpha .
\] 
The corresponding pressure is
\be
P_C = - m c^2 \left( \frac{m c}{\hbar} \right)^3 \frac{9}{4 \pi}
\frac{Z^{2/3} \alpha^5}{10} \frac{1}{r_{e}^{4}} .
\ee
In this model, the total pressure is then given by 
\be
P = P_e + P_C .
\ee
Thus, since $P_C$ is negative, the equation of state would
lead to negative pressures at low density (\cite{salpeter}).  

In integrating the equations of hydrostatic equilibrium, 
we follow the classical procedure of Hamada \& Salpeter (1961). 
Figures 1, 3, and 5 show the mass-radius relation of $^{4}$He, $^{12}$C, 
and $^{56}$Fe white dwarfs for a given magnetic field strength.
Figures 2, 4, and 6 show the relation between mass and central density
of these white dwarfs for a given magnetic field strength.
Here it can be seen that our results approach the Hamada \& Salpeter (1961) results 
as the magnetic field strength decreases. 
For high central field strengths $\gm \simeq 0.01 - 1$ 
[$B \simeq 4.4 \times (10^{11} - 10^{13})$ G], 
both the mass and radius of magnetic white dwarfs increase compared to non-magnetic 
white dwarfs of the same central density. 
For instance, for $\gm \approx 0.8$ carbon white dwarfs, the radius
increases by about 30\% for $M \approx 1 \, M_{\odot}$. 
Similarly for $R \approx 0.01 R_{\odot}$ the mass also increases by about 25\%. 
These results are approximately consistent with Ostriker \& Hartwick (1968), i.e.
the radius increases while the central density $\rho_c$ decreases 
as the magnetic field increases for fixed $M$. 
As expected, for $B \lsim 10^{10}$ G, 
internal magnetic fields do not affect the white-dwarf mass-radius relation.
\vspace*{1.5cm}
\placefigure{fig1}
\begin{center}
\vspace*{0.5cm}
{\epsfxsize=6.5cm 
\epsfbox{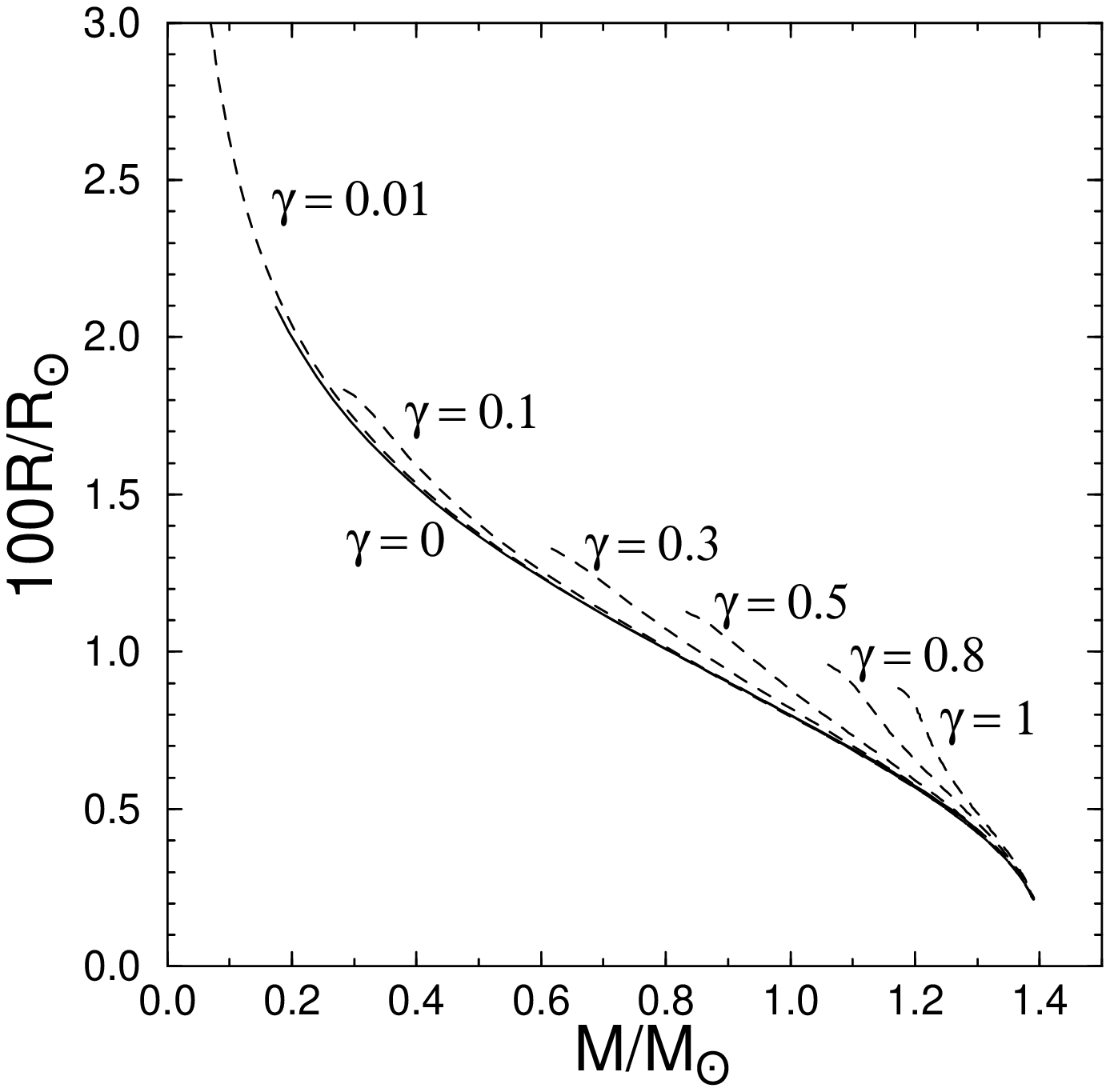}}
\vskip 1.0cm
\end{center}
{\small {\sc Fig. 1}
Relation between the mass $M$ and radius $R$ of a $^{4}$He magnetic white dwarf
for the indicated magnetic-field strengths. The solid line denotes the Hamada \& Salpeter
model for non-magnetic white dwarfs ($\gm = 0$).
The dashed lines are magnetic white dwarfs.}
\vskip 1.5cm
\placefigure{fig2}
\begin{center}
{\epsfxsize=6.5cm 
\epsfbox{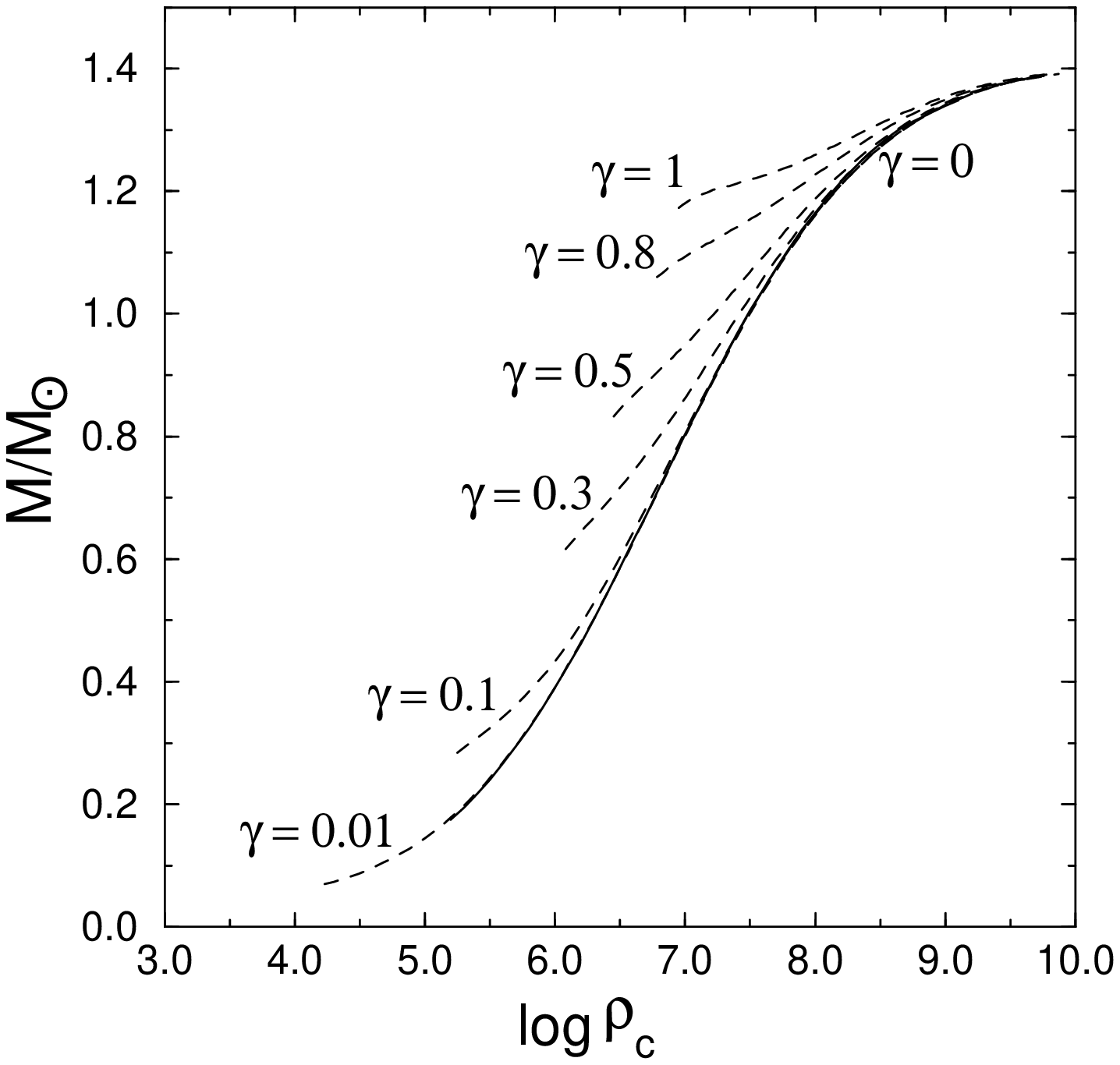}}
\vskip 1.0cm
\end{center}
{\small {\sc Fig. 2}
Relation between the central density $\rho_c$ (in g/cm$^3$) and mass $M$
for $^{4}$He magnetic white dwarfs. The solid line denotes the Hamada-Salpeter
model for non-magnetic white dwarfs ($\gm = 0$).  
The dashed lines are for magnetic white dwarfs.}
\vskip 1cm
A striking feature of these results is that white dwarfs with strong interior magnetic
fields should be massive. 
This is simply because a star becomes unbound if the magnetic plus matter pressure force
exceeds the gravitational force.
For example, there is no stable solution with
$M \lsim 0.5 M_{\odot}$ for $\gm = 0.3$ carbon white dwarfs.
This is consistent with recent observations that on average magnetic white dwarfs have a
higher mass than typical non-magnetic white dwarfs (see Table 1).
Note, however, that for a field strength of $ B \lsim 1 \times 10^{13}$ G, 
magnetic fields only give a relatively small change on the mass-radius relation, that is, 
we can not distinguish between magnetic and non-magnetic white dwarfs for
$\gamma \lsim 0.2$. 

\section{Discussion}

In this work, we have calculated the equation of state for an electron gas in a 
magnetic field at zero temperature. For simplicity, we have assumed a uniform 
composition to obtain the relation between mass and radius for magnetic white dwarfs.
For high internal magnetic fields 
$B \simeq 4.4 \times (10^{11} - 10^{13})$ G, ($\gm \simeq 0.01 - 1$),
the mass-radius relation is modified.
Our results not only confirm the  Ostriker \& Hartwick (1968) result of increasing
radius and decreasing central density $\rho_c$ with increasing field, but also are
consistent with the suggestion (Liebert 1988) that observed magnetic white
dwarfs have masses which are on average larger than non-magnetic white dwarfs,
implying more massive and younger progenitors.

The question remains, however, as to whether it is reasonable to consider such high 
internal field strengths for magnetic white dwarfs.
First, one must assume that the magnetic fields are well hidden beneath the
surface, while the surface fields are several orders of magnitude less. 
Second, assuming that flux is conserved during the collapse to a white dwarf, 
the progenitor of the white dwarfs must have had sufficiently large fields
to produce the required white dwarf internal field strengths.
Flux conservation implies that the central field strength of the progenitors 
is of order $\sim 10^8$ G, assuming $R \approx 1 \, R_{\odot}$.
This is reasonable. 
During star formation, the collapse of a typical interstellar cloud with
radius $\sim 0.1$ pc, mass $\sim 1 M_{\odot}$, and protostellar magnetic field of
magnitude $\sim 3 \times 10^{-6}$ G would result in a field strength of 
$\sim 3 \times 10^8$ G
in a solar type star formed from this material (\cite{spitzer}).
Although there is no evidence for main sequence stars with such field strengths,
they are not ruled out by observations either (\cite{ST}).

\section{Comparison with observations}

Figure 7 shows a comparison between our calculations and white dwarfs with known masses
and radii from the HIPPARCOS survey (\cite{vauclair}; \cite{provencal}).
Plotted error bars are the quoted $\pm 1 \sigma$ observational uncertainties.
Strong hidden interior magnetic fields would be expected to manifest themselves by a
preponderance of stars with large masses and radii.
Most of the data, however, are within $2 \sigma$ of the non-magnetic theoretical curves.
A puzzling feature however (\cite{provencal}) is that some of the best determined data
points (e.g., EG 50 and Procyon B) can not be fit without postulating an iron composition,
something which seems unlikely from a stellar evolution standpoint.
Furthermore, although the evidence is not compelling, there are at least
two well determined stars [i.e., GD 140 and Sirius B (\cite{provencal})]
as well as some field stars with $M \gsim 0.6 M_{\odot}$ and
$R \gsim 0.012 R_{\odot}$ which may be better fit if strong internal magnetic fields
are assumed. However, Sirius B and some of the field stars can also be accommodated
by atmospheric models (\cite{wood}).

\vspace*{1.5cm}
\placefigure{fig3}
\begin{center}
\vspace*{1.5cm}
{\epsfxsize=6.5cm
\epsfbox{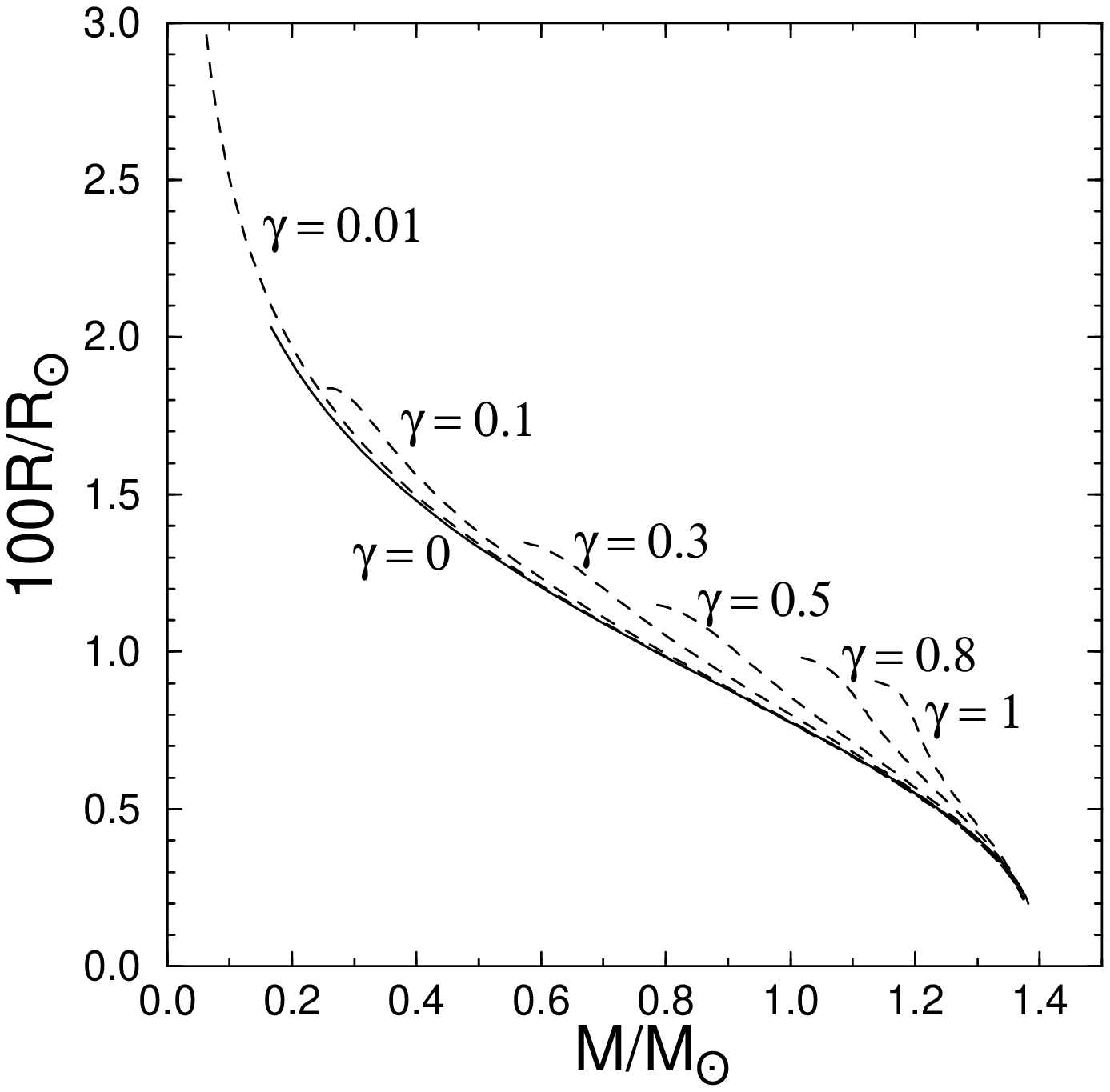}}
\vskip 0.5cm
\end{center}
{\small {\sc Fig. 3}
Same as Fig. 1, but for $^{12}C$.}
\placefigure{fig4}
\begin{center}
\vspace*{1.5cm}
{\epsfxsize=6.5cm
\epsfbox{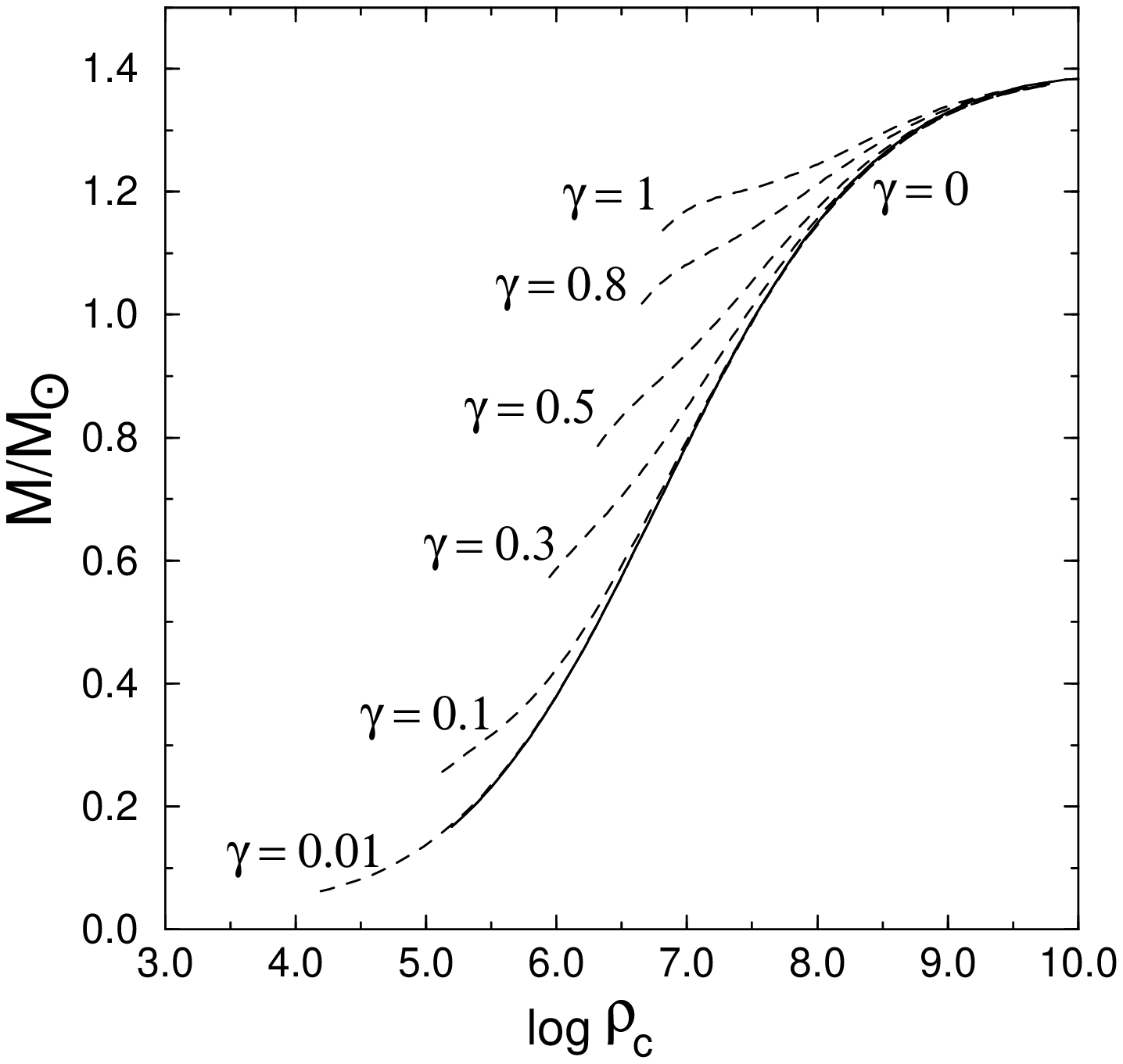}}
\vskip 0.5cm
\end{center}
{\small {\sc Fig. 4}
Same as Fig. 2, but for $^{12}C$.}
\vspace*{1.5cm}
\placefigure{fig5}
\begin{center}
\vspace*{1.5cm}
{\epsfxsize=6.5cm
\epsfbox{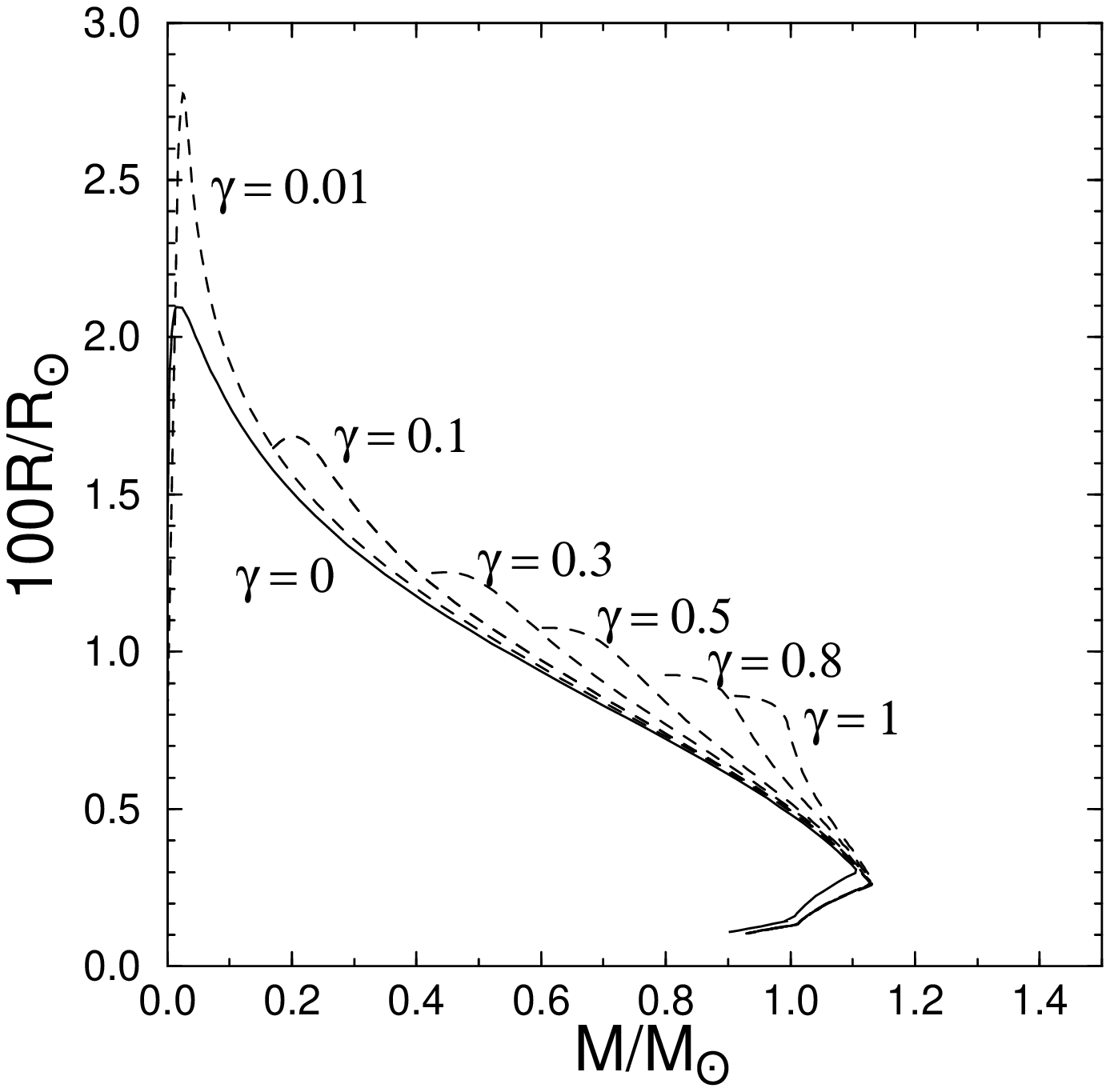}}
\vskip 0.5cm
\end{center}
{\small {\sc Fig. 5}
Same as Fig. 1, but for $^{56}Fe$.}
\placefigure{fig6}
\begin{center}
\vspace*{1.5cm}
{\epsfxsize=6.5cm
\epsfbox{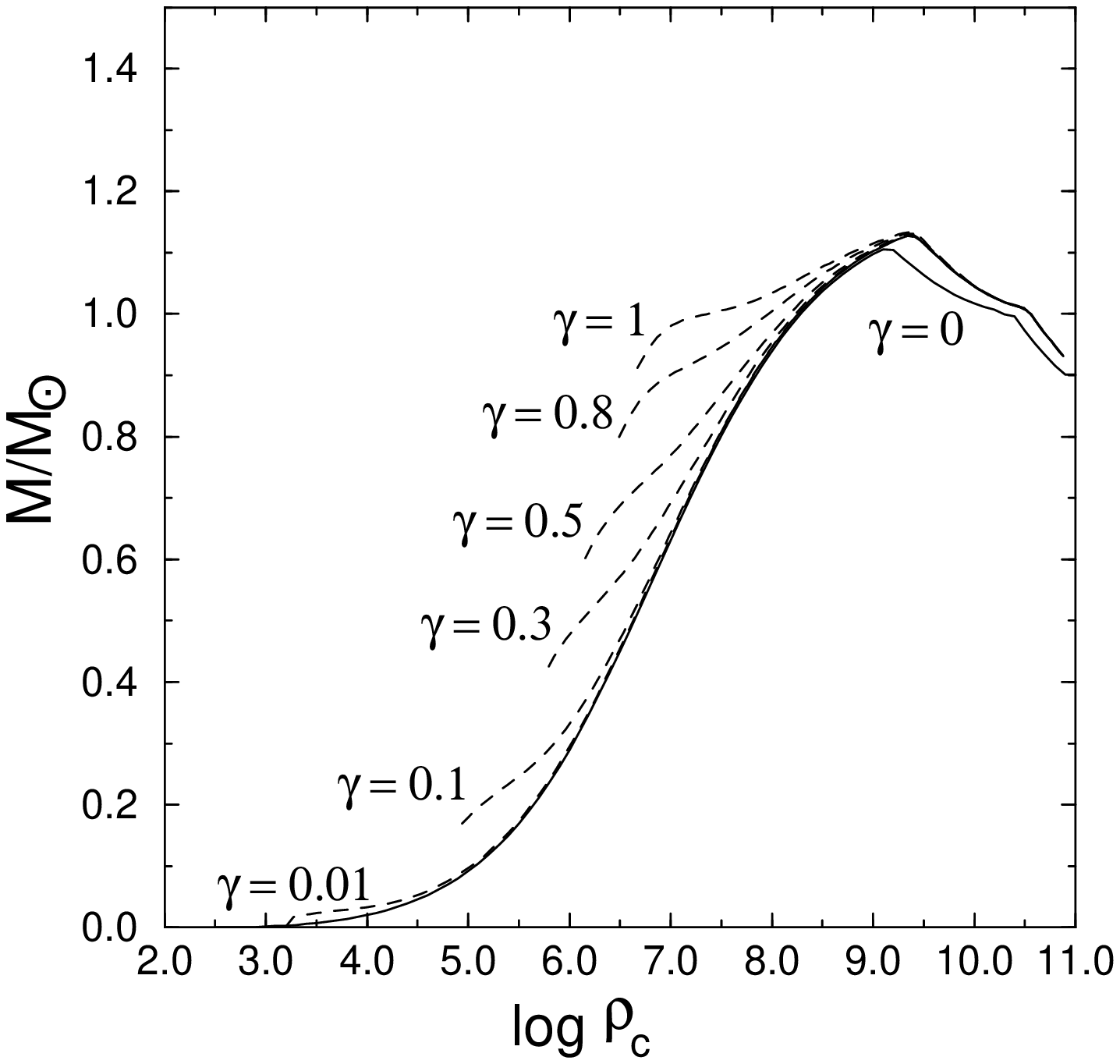}}
\vskip 0.5cm
\end{center}
{\small {\sc Fig. 6}
Same as Fig. 2, but for $^{56}Fe$.}
\placefigure{fig7}
\vspace*{1.5cm}
\begin{center}
\vspace*{0.5cm}
{\epsfxsize=6.5cm
\epsfbox{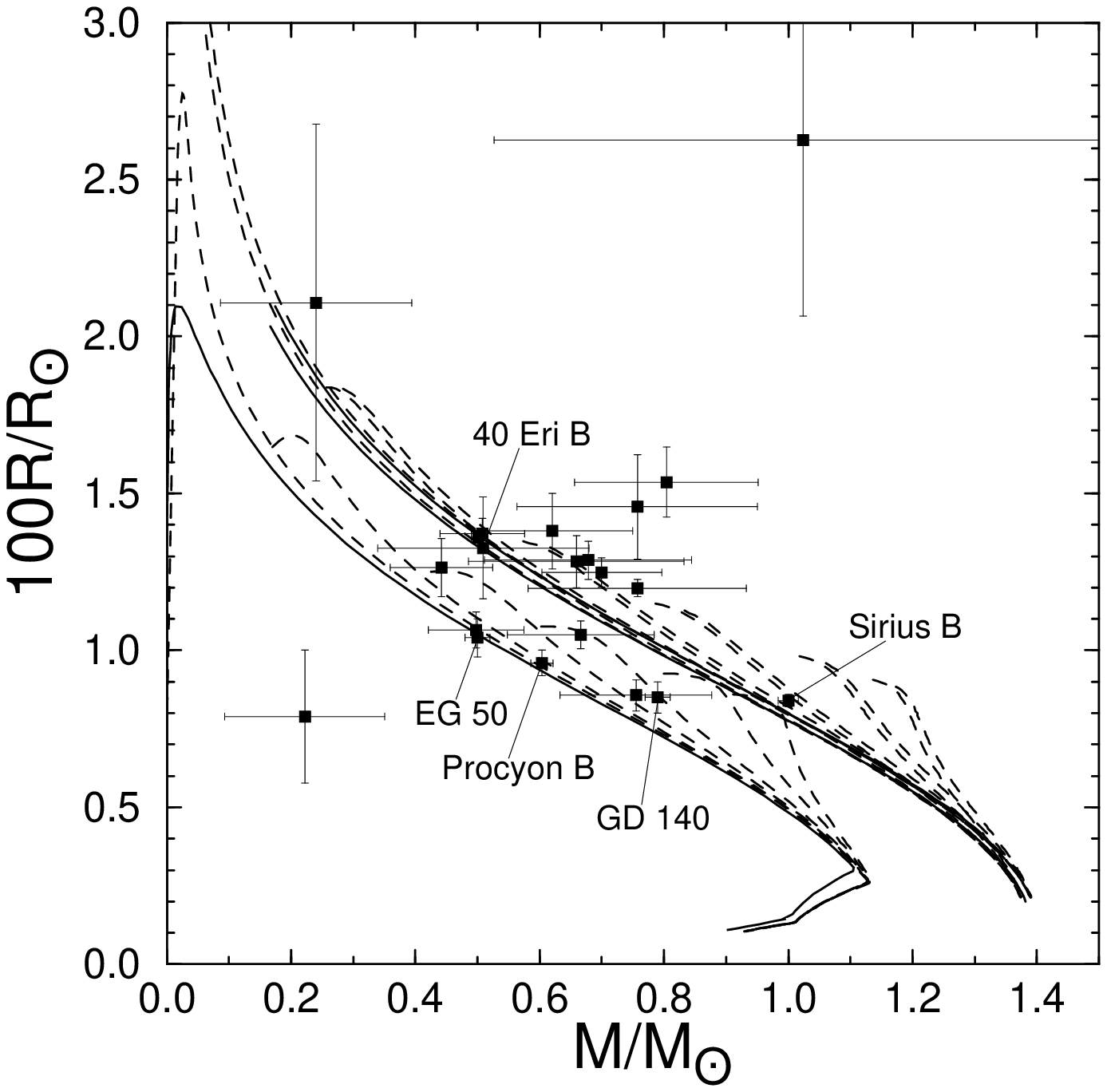}}
\vskip 0.5cm  
\end{center}
{\small {\sc Fig. 6}
A comparison of white dwarfs with known masses and radii from the HIPPARCOS survey
at the $1 \, \sigma$ level (\cite{vauclair}; \cite{provencal})
with our calculations from Figs. 1, 3, and 5. The upper curves are for $^{4}$He and
$^{12}$C white dwarfs. The lower curves are for $^{56}$Fe white dwarfs.}
\vskip 1.cm
Some typical magnetic white dwarfs which have reported masses $M$ and
surface magnetic fields $B_s$ are summarized in Table 1.
This table shows that magnetic white dwarfs are typically more massive than 
non-magnetic white dwarfs (on average $M \sim 0.6 M_{\odot}$).
Among them RE J0317-858 ($B_s > 10^{8}$ G) is approaching the Chandrasekhar mass
limit (\cite{barstow}).

Not listed in Table 1, however, are the two known magnetic white dwarfs with 
the strongest surface magnetic field,
GD 229 ($B_s \gsim 10^9 \,$ G) and PG 1031+234 ($B_s \approx (0.5 - 1) \times 10^9$ G).
Unfortunately, there are no reported masses or radii for those 
magnetic white dwarfs. If their masses and radii could be measured, 
then our magnetic field model might be tested.  

Perhaps the most interesting object in Table 1 is LB 11146 (PG 0945+245). This is 
an unresolved binary system consisting of two degenerate stars (\cite{liebert93}):
one component is a normal DA white dwarf ($M = 0.91 \pm 0.07 \, M_{\odot}$) 
with no detectable magnetic field; the other has a strong magnetic field
($B_s > 3 \times 10^8$ G) and a mass in the range  
$0.76 \le M/M_{\odot} \lsim 1.0$.
Thus, assuming for illustration a strong interior field of $\gamma \simeq 0.5$,
and that this is a typical equal mass system with $M = 0.9 \, M_{\odot}$ each,
then the radius of the magnetic component would be larger by about 10\% than 
the normal star. Alternatively, if they have similar radii of $R = 0.01 \, R_{\odot}$, 
the magnetic star would have a heavier mass by about 12\% than the normal one 
for $\gm \simeq 0.5$. 

Regarding the radius of magnetic white dwarfs, Greenstein \& Oke (1982) have reported
radii of $R \sim 0.0066 \, R_{\odot}$ for Grw +70$^\circ$8247
($B_s \approx 3.2 \times 10^8$ G) and $R \sim 0.01 \, R_{\odot}$ for Feige 7
($B_s \approx 3.5 \times 10^7$ G) from the interpretation of their spectra.
Assuming an interior field which is about $10^5$ times stronger than the surface field,
then we would deduce $\gm \approx 0.8$ and $M \sim 1.1 - 1.2 \, M_{\odot}$
for Grw +70$^\circ$8247,
consistent with the expectation of high masses for magnetic white dwarfs.
Similarly, we would deduce $\gm \approx 0.075$ and $M \sim 0.8 \, M_{\odot}$ for Feige 7.
This star however would be nearly indistinguishable from a non-magnetic white dwarf.

\placetable{tbl-1}
\begin{table*}
\caption{Mass and surface magnetic field strength in some typical magnetic white dwarfs}
\begin{center}
\begin{tabular}{|c|c|c|c|} \hline
                      & $M (M_{\odot})$ & $B_s (10^6$ G) & Ref. \\ \hline
   PG 2329+267        &   $\sim$ 0.9    &  2.3           & Moran et al. 1998 \\ \hline
    LB 11146B         &   0.76 - 1.0    &  $>$ 300       & Liebert et al. 1993 \\ \hline
 Grw +70$^{\circ}$8247&   $>$ 1.0       &  320           & Greenstein \& Oke 1990 \\ \hline 
 1RXS J0823.6-2525    &     1.2         & $\sim 3$       & Ferrario et al. 1998 \\ \hline
   PG 0136+251        &     1.28        & 1.3$\,?$       & Vennes  et al. 1997   \\ \hline
   PG 1658+441        &     1.31        & 3.5$\,?$       & Schmidt et al.  1992\\ \hline
 RE J0317-858         &     1.35        &  660           & Barstow et al. 1995\\ \hline
\end{tabular}
\end{center}
\end{table*}

\acknowledgments

This work supported in part by NSF Grant-97-22086 and DOE Nuclear Theory Grant
DE-FG02-95ER40934.


\end{document}